\documentclass[a4paper,fleqn]{cas-dc}

\usepackage[numbers]{natbib}
\usepackage{rotating}
\usepackage{multirow}
\usepackage{xcolor}
\usepackage{color}
\usepackage{booktabs}
\usepackage{svg}

\def\tsc#1{\csdef{#1}{\textsc{\lowercase{#1}}\xspace}}
\tsc{WGM}
\tsc{QE}
\tsc{EP}
\tsc{PMS}
\tsc{BEC}
\tsc{DE}
\begin{document}
\let\WriteBookmarks\relax
\def\floatpagepagefraction{1}
\def\textpagefraction{.001}

\shorttitle{Leveraging GANs for IDS Enhancement}

\shortauthors{}

\title [mode = title]{Enhancing Network Intrusion Detection   Performance using Generative Adversarial Networks }

%\tnotemark[1,2]

%\tnotetext[1]{This document is the results of the research
 %  project funded by the National Science Foundation.}

%\tnotetext[2]{The second title footnote which is a longer text matter
 %  to fill through the whole text width and overflow into
 %  another line in the footnotes area of the first page.}

\author[]{Xinxing Zhao}
[type=editor,
 orcid=0000-0001-5815-043X]

\cormark[1]


\ead{xinxing.zhao@stengg.com}

%\ead[url]{www.cvr.cc, cvr@sayahna.org}

%\credit{Conceptualization of this study, Methodology, Software}

%\affiliation[1]
\affiliation[]{organization={ST Engineering},
    country={Singapore.}} %Cybersecurity Strategic Technology Centre},
   % addressline={600 West Camp Road}, 
    % citysep={}, % Uncomment if no comma needed between city and postcode
%    postcode={797796}, 
%    country={Singapore}}

\author[]{Kar Wai Fok}
[  ]

 \ead{fok.karwai@stengg.com}

\author[]{Vrizlynn L. L. Thing}[%
   ]
\ead{vriz@ieee.org}

%\credit{Data curation, Writing - Original draft preparation}

\cortext[cor1]{Corresponding author}
%\cortext[cor2]{Principal corresponding author}

%\fntext[fn1]{This is the first author footnote. but is common to third
%  author as well.}
%\fntext[fn2]{Another author footnote, this is a very long footnote and
%  it should be a really long footnote. But this footnote is not yet
%  sufficiently long enough to make two lines of footnote text.}

%\nonumnote{This note has no numbers. In this work we demonstrate $a_b$
%  the formation Y\_1 of a new type of polariton on the interface
%  between a cuprous oxide slab and a polystyrene micro-sphere placed
%  on the slab.
  %}

\begin{abstract}
Network intrusion detection systems (NIDS) play a pivotal role in safeguarding critical digital infrastructures against cyber threats. Machine learning-based detection models applied in NIDS are prevalent today. However, the effectiveness of these machine learning-based  models is often limited by the evolving and sophisticated nature of intrusion techniques as well as the lack of diverse and updated training samples. In this research, a novel approach for enhancing the performance of an NIDS through the integration of Generative Adversarial Networks (GANs) is proposed. By harnessing the power of GANs in generating synthetic network traffic data that closely mimics real-world network behavior, we address a key challenge associated with NIDS training datasets, which is the data scarcity.  Three distinct GAN models (Vanilla GAN, Wasserstein GAN and Conditional Tabular GAN)  are implemented in this work to generate authentic network traffic patterns specifically tailored to represent the anomalous activity. We demonstrate how this synthetic data resampling technique can significantly improve the performance of the NIDS model for detecting such activity.  By conducting comprehensive experiments using the CIC-IDS2017 benchmark dataset, augmented with GAN-generated data, we offer empirical evidence that shows the effectiveness of our proposed approach. Our findings show that the integration of GANs into NIDS can lead to enhancements in intrusion detection performance for attacks with limited training data, making it a promising avenue for bolstering the cybersecurity posture of organizations in an increasingly interconnected and vulnerable digital landscape.
\end{abstract}

% \includegraphics{figs/grabs.pdf}


%\item Research highlights item 1
%\item Research highlights item 2
%\item Research highlights item 3

% Keywords

\maketitle

\section{Introduction}
\indent The proliferation of devices and the ever-expanding Internet of Things (IoT) ecosystem nowadays have led to an unprecedented level of connectivity. From smartphones and laptops to smart refrigerators and industrial machinery, an increasingly diverse array of devices are constantly generating massive amounts of data. While this interconnectedness has ushered in remarkable convenience and innovation, it has also exposed us to a growing threat: malicious network intrusion.  The more devices we connect to networks, the larger the attack surface becomes \cite{hussain2020deep}. Attackers are quick to capitalize on this expanded attack surface, targeting vulnerabilities in both traditional and IoT devices \cite{gupta2022smart, kampourakis2023systematic}. 

These malicious actors launch various types of attacks, such as botnet traffic  \cite{mustapha2023detecting}, malware infections \cite{beaman2021ransomware}, zero-day exploits \cite{zhao2018multi}, man-in-the-middle attacks  \cite{conti2016survey}. NIDS are specialized tools designed to monitor network traffic, analyze data packets, and identify any suspicious or unauthorized activities, and they play a pivotal role in safeguarding networks against evolving threats.  The widespread integration of machine learning models into NIDS \cite{ahmad2021network} is evident today and brings forth many capabilities. These models empower NIDS to deliver real-time \cite{vishwakarma2022dids}, adaptive \cite{chindove2021adaptive}, and exceptionally effective \cite{injadat2020multi} threat detection, therefore serving as a defence mechanism to shield the more vulnerable devices within the  interconnected landscape and fortify the security of the invaluable data.

In addition to the well-recognized challenges that NIDS face, such as handling encrypted traffic \cite{wang2023feature}, combating advanced evasion techniques \cite{afianian2019malware}, addressing scalability concerns \cite{raja2020sp}, and navigating the complexities of network traffic patterns \cite{yang2021mth}, there are still significant obstacles that exist  during the NIDS training process, such as sample scarcity and class imbalance \cite{binbusayyis2021unsupervised}. Sample scarcity refers to the scarcity of relevant data, particularly when it comes to capturing anomalous network traffic indicative of cyberattacks. This scarcity is rooted in the fact that real-world network attacks occur less frequently in comparison to the vast volume of normal network traffic. Consequently, the task of collecting a sufficiently robust dataset for training an effective intrusion detection model becomes a  challenge.  Class imbalance arises because normal network traffic, which is abundant in most datasets, significantly outweighs the occurrence of attack instances. This  disproportion can be problematic, as machine learning algorithms tend to exhibit a bias towards the majority class when faced with imbalanced datasets. This bias can lead to NIDS models being less sensitive to detecting network attacks since they predominantly focus on the abundant normal traffic class.

Existing literature offers several strategies to deal with the challenges posed by sample scarcity and class imbalance, such as data augmentation techniques \cite{liu2022intrusion}, resampling methods \cite{lee2021gan}, ensemble learning approaches \cite{tama2021ensemble}, and sophisticated feature engineering \cite{thakkar2022survey}. In this research, we focus on the problem of data scarcity and propose a method leveraging the capabilities of generative AI, and, in particular, GAN models to generate attack samples in enhancing the NIDS detection performance based on the CIC-IDS2017 dataset \cite{sharafaldin2018toward}. 

The primary contributions of the research are outlined below:  
Firstly, this study addresses a critical challenge within intrusion detection systems: the limited availability of representative attack samples for training IDS. To overcome this challenge, three distinct GAN models are employed to generate additional attack samples, based on the relatively new CIC-IDS2017 dataset. Secondly, different from the majority of the related studies in the literature, several mechanisms have been applied in assessing the quality of the generated attack samples. 
To ensure confidence in the GAN models and underscores their reliability, we have employed metrics and methodologies to assess the closeness and likeness between the initial samples and the subsequently generated ones. Thirdly, in contrast to previous studies, this study explores the generation of varying quantities of attack samples and their subsequent integration into the original dataset. Through comprehensive experimentation and rigorous testing on these augmented datasets, it is  demonstrated that with more data samples generated, improvements in intrusion detection system performance can be achieved. Notably, our approach outperforms the existing advanced techniques in terms of precision, recall, and F1-score metrics. 
%\textcolor{blue}{
The rest of this paper is organized in the following manner. Section \ref{Related work} describes GANs-related  research on network intrusion. Section \ref{Methodology}  introduces the methodology and three GAN models applied in this research. Section \ref{enhancing} describes the experiments and performance enhancement by these generated attack samples. Section \ref{Discussion} provides the discussions and section \ref{Conclusion}  presents the conclusions.

\section{Related Work} \label{Related work}

There are three primary types of network intrusion detection systems. The first type comprises signature-based systems \cite{hubballi2014false}, they evaluate incoming traffic against a pre-established database containing recognized attack patterns. If a match is found, an alert will be triggered. This method is effective in identifying known threats, however, may struggle with novel attacks \cite{khraisat2019survey}.

The second type of  systems is anomaly-based \cite{jabez2015intrusion}. Anomaly-based NIDS establishes a baseline of normal network behavior by continuously monitoring network traffic and system activities and comparing them to this baseline. When any deviation or anomaly is detected, an alert will be triggered. This detection approach is highly effective in recognizing novel or previously unencountered threats that lack established signatures. However, it does come with certain drawbacks, including the need for complex configurations and substantial computational resource demands.  One highly related and complementary aspect to anomaly-based detection in NIDS is classification \cite{ravipati2019intrusion}. Anomaly detection focuses on identifying outliers or suspicious patterns, whereas classification specializes in categorizing the identified anomalies into specific threat types, and once anomalies are detected, classification models can be applied to classify the nature of the threat, providing more detailed insights into the specific attack type. Therefore,  the fusion of these two elements can improve the overall performance of the system and provides more actionable information to security teams. And lastly, hybrid systems \cite{dalai2017hybrid}, combining the strengths of both previous mentioned two approaches, are more adaptable and offer a balanced approach to threat detection. They increase the accuracy of identifying both known and unknown threats.

 Generative Adversarial Networks \cite{goodfellow2014generative}, a relatively recent advancement in the realm of machine learning, are applied in unsupervised learning tasks that primarily focused on generating  data that resembles a given dataset.  They have found applications in many fields and industries such as  computer vision \cite{ozkanouglu2022infragan}, natural language processing \cite{liang2021text}, generative art \cite{civit2022systematic}, and anomaly detection\cite{schlegl2019f}. GAN models can  be integrated into these intrusion detection systems to enhance their capabilities, as they can generate synthetic data for training \cite{shahriar2020g}, improve feature extraction \cite{zhu2022black}, and address some of the challenges associated with both signature-based and anomaly-based methods. For instance, GANs offer the capability to produce varied synthetic attack data. This empowers signature-based systems to adjust to emerging attack patterns, thereby enhancing their accuracy in identifying variations of known attacks. GANs can also create synthetic normal traffic data, which can be used to enrich the baseline of normal behavior in anomaly-based systems. 
 
GANs also have been utilized directly as network intrusion detection mechanisms. For instance, Patil et al. \cite{patil2022network}  introduced a framework that leverages a bidirectional GAN for anomaly detection. This framework was assessed using the KDDCUP-99 dataset \cite{tavallaee2009detailed} and was subsequently compared with  other deep learning models to evaluate its performance. Truong et al. \cite{truong2020empirical} adopted two GAN models with newly developed neural networks for generators and discriminators. They carried out extensive experiments based on  datasets such as CIC-IDS 2017 and UNSW-NB15 \cite{moustafa2015unsw} to assess the effectiveness of GANs, comparing them with established unsupervised detection methods.

 GAN models are also  increasingly employed to generate network traffic that closely mimics normal patterns but is, in fact, malicious in nature, with the intent of evading intrusion detection systems. For instance, Chauhan et al. \cite{chauhan2020polymorphic}
  introduced a GAN model for generating synthetic DDoS traffic. This model dynamically alters the number of attack features and swaps features with unused attack features in the training set. As a result, the attacks they generated can effectively evade detection by the IDS. Lin et al. \cite{lin2022idsgan}  introduced a framework based on GANs that possesses the ability to transform original malicious network traffic into traffic that mimics normal behavior while preserving the attack functionalities.This framework dynamically learns the functioning of a real-time black-box detection system and employs the modified attack traffic to effectively evade such detection systems. In another study, Mustapha et al. \cite{mustapha2023detecting} employed a GAN model to align DDoS functional features with the distribution of benign sample features, effectively perturbing the data. Their conclusion highlighted that the introduction of perturbations in the input features significantly decreased the performance of the IDS.

 GANs have also demonstrated their applicability in NIDS as a means to enhance their performance and robustness. For instance, Lee et al.  \cite{lee2021gan}  applied GANs to generate network traffic, specifically targeting the mitigation of the class imbalance problem commonly encountered in NIDS datasets. Their research findings favored GANs over conventional techniques like SMOTE \cite{zhang2022iot}. Shahriar et al. \cite{shahriar2020g} employed an ANN-based GAN model for generating synthetic samples. They then trained an IDS on both the synthetic samples and the original ones, utilizing the NSL-KDD dataset\cite{tabassum2022fedgan}. Their investigation revealed that the IDS model incorporating GAN functions outperforms a standalone IDS significantly in detecting attacks.
Bourou et al. \cite{bourou2021review}  used  CTGAN \cite{xuctgan2019}, CopulaGAN \cite{upadhyay2023comparative}, and TableGAN \cite{park2018data} models to generate synthetic DoS attacks using the NSL-KDD datase and concluded that the generated datasets are suitable for training various machine learning models.  In this ongoing research, the primary focus lies in harnessing the power of GAN models to improve the classification effectiveness of an IDS. This approach involves the generation of larger sets of attack samples, which are subsequently employed to train the IDS.  The current research distinguishes itself from other related and recent studies in the literature such as \cite{lee2021gan, shahriar2020g, bourou2021review} in three significant ways.  First,  a relatively new and highly advantageous dataset, CIC-IDS2017 is leveraged. This dataset offers several key benefits, including the inclusion of realistic network traffic, a substantial volume of network data, and a wide spectrum of attack scenarios, aligning closely with real-world conditions.  Second, this research utilizes multiple GAN models for generating specific attack samples. These generated attack samples are further being assessed for their closeness to the original samples, subjecting them to various testing and comparison methodologies. Notably, the newly generated samples are employed independently for training the IDS, and their impact on enhancing IDS classification is evaluated. Third, an exploration is conducted into the generation of diverse quantities of attack samples, which are then integrated separately into the CIC-IDS2017 dataset. This approach facilitates the observation of scalability in IDS performance enhancements across different sample sizes.

The adoption of this threefold approach distinguishes the current research and contributes to a more comprehensive understanding of intrusion detection system performance within the context of GAN-generated attack samples. By broadening the scope of attack samples through GAN-generated data, the aim is to create a more robust and adaptive IDS, enhancing its capability to safeguard network integrity and security against evolving and sophisticated intrusion attempts.

\section{Methodology}\label{Methodology}

\begin{figure}[!ht]
  \includegraphics[width=1.0 \linewidth]{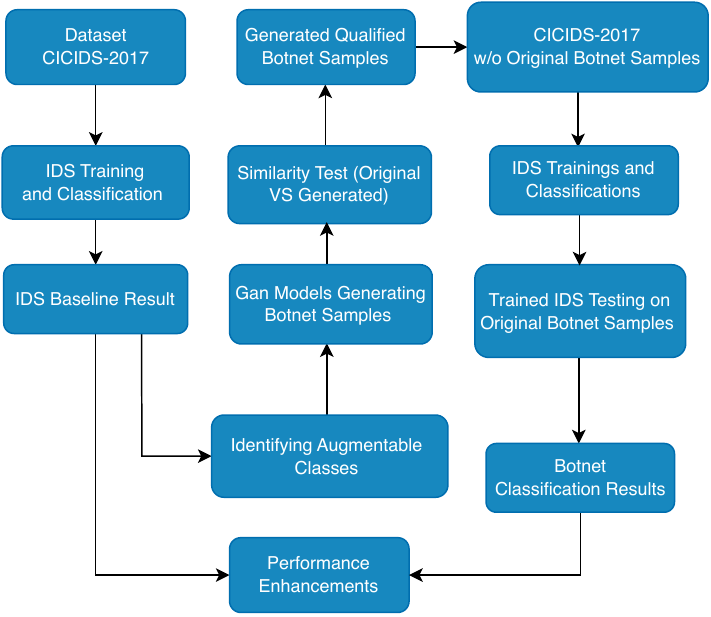}
  \caption{The Flow of Enhancing NIDS Performance with GANs }
  \label{fig:flow}
\end{figure}

  We start by introducing the CIC-IDS2017 dataset, outlining data processing methods, and categorizing (regrouping) it into broader classes. Following this, we present an IDS designed to classify these general classes, thereby establishing a baseline for the IDS's performance. We then delve into the fundamentals of GANs and describe three GAN models that we've implemented in this study. Subsequently, we elaborate on our approach to handling the Botnet samples within the CIC-IDS2017 dataset and explain how we generate new Botnet samples based on the original dataset. Then, we outline the methodology employed to assess the similarity between the generated and the original samples. We subsequently integrate the newly generated Botnet samples into the original dataset and observe the resulting performance enhancements of the IDS. Figure \ref{fig:flow} illustrates the entire process of enhancing the NIDS with additional attack samples generated using GANs. %  More details  regarding the incorporation of newly generated Botnet samples into the original dataset and the contributes to the performance improvements of the IDS will be presented in Section \ref{enhancing}.

\subsection{The Dataset}

The dataset employed in this research, namely the CIC-IDS2017, is a publicly accessible dataset specifically designed for evaluating and benchmarking intrusion detection systems  and network security solutions.

We have chosen the CIC-IDS2017 dataset for our research experiments for several compelling reasons. Firstly, this dataset is relatively new and encompasses a wide spectrum of network activities, including various types of attacks and benign traffic. Given that our research primarily revolves around generating a specific class of network attack samples to enhance the classification performance of a multi-class IDS, CIC-IDS2017 proves to be an ideal choice. Secondly, CIC-IDS2017 has been available for a considerable amount of time and has remained unchanged, ensuring stability and consistency. This unchanging nature of the dataset serves as a reliable reference point for our research. Lastly, CIC-IDS2017 has gained widespread acceptance within the research community. This widespread adoption facilitates easy comparison of our results with the work of other researchers, fostering knowledge sharing in the field. 
\begin{table}
    \centering
    \caption{Processed CICIDS-2017 dataset.}
    \label{tab:CICIDS2017}
    \begin{tabular}{|c|c|}
        \hline
        Classes & Instances \\
        \hline
        Benign & 2271320 \\
         \hline
        DoS Hulk & 230124 \\
         \hline
        PortScan & 158804 \\
         \hline
        DDoS & 128025 \\
         \hline
        DoS GoldenEye& 10293 \\
         \hline
        FTP-Patator & 7935 \\
         \hline
        SSH-Patator & 5897 \\
         \hline
        DoS slowloris & 5796 \\
         \hline
        DoS Slowhttptest & 5499 \\
         \hline
        Bot & 1956\\
         \hline
        Web Attack: Brute Force & 1507 \\
         \hline
        Web Attack: XSS & 652 \\
         \hline
       Infiltration & 36 \\
        \hline
        Web Attack:  Sql Injection  & 21\\
         \hline
        Heartbleed   & 11 \\

        \hline
    \end{tabular}
\end{table}

After checking for and handling null values and infinity values in the dataset - dropping any rows that contained NaN, Null or Inf values - different class of network activities and number of their instances are presented  in Table   \ref{tab:CICIDS2017}.

\begin{table}
    \centering
    \caption{Grouping of original classes into more general classes.}
    \label{tab:classgroup}
    \begin{tabular}{|c|c|}
        \hline
        New Classes & Original Classes \\
        \hline
        Benign & Benign \\
        \hline
        Botnet & Bot \\
        \hline
        \multirow{2}{*}{Brute Force} & FTP-Patator \\
        & SSH-Patator \\
        \hline
        DDoS & DDoS \\
        \hline

         \multirow{5}{*}{DoS} & DoS GoldenEye \\
        & DoS Hulk \\
        & DoS Slowhttptest \\
        & DoS Slowloris \\
        & Heartbleed \\
        \hline
        
        Probe & ProtScan \\
        \hline
        \multirow{3}{*}{Web Attack} & Web Attack: Brute Force \\
        &  Web Attack: SQL Injection \\
        & Web Attack: XSS \\
        \hline
        Infiltration & Infiltration \\
        \hline
    \end{tabular}
\end{table}

To reduce the class imbalance issue within this dataset, we adopted a similar approach as a prior study related to CIC-IDS2017 \cite{panigrahi2018detailed}. This involved the re-labeling and grouping of classes to create new (more general) classes. The reasoning behind this choice arises from the observation that certain attack types share strong similarities, such as Dos GoldenEye, Dos Hulk, DoS Slowloris, and DoS Slowhttptest, which we grouped into a unified "Dos" class. Additionally, even after this regrouping, we still had eight distinct classes available for further researching in our study.

Table \ref{tab:classgroup} shows how to form new classes based on the original classes while  Table \ref{tab:newclass} shows the new classes and their number of instances.

\begin{table}
    \centering
    \caption{New classes and their instances.}
    \label{tab:newclass}
    \begin{tabular}{|c|c|}
        \hline
        New Classes & Number of Instances \\
        \hline
        Benign & 2271320 \\
         \hline
        DoS & 251723 \\
         \hline
        Probe & 158804 \\
         \hline
        DDoS & 128025 \\
         \hline
        Brute Force& 13832 \\
         \hline
    Web Attack& 2180\\
     \hline
        Botnet & 1956\\
         \hline
       Infiltration & 36 \\
        \hline
    \end{tabular}
\end{table}

\subsection{The IDS Baseline and Motivation} \label{sec:Baseline}

Previous research \cite{priyanka2022performance,reis2019selection, stiawan2020cicids} has established that the utilization of Random Forest (RF) models as classifiers can result in robust classification performance on the CIC-IDS2017 dataset. In this study, a RF model is  adopted as well,  for classifying the newly formed  multiple classes. The  chi-squared (chi2) statistical test is employed as the scoring function to select the top 32 features. In this baseline (the ratio between the training set and testing set is 8 to 2), the RF model reaches a classification accuracy of 0.9972, which aligns closely with the findings of a prior investigation \cite{stiawan2020cicids}.

\begin{table}[htbp]
\centering
\caption{Baseline IDS Classification performance.}
\begin{tabular}{|l|c|c|c|}
\hline
\textbf{New Classes} & \textbf{Precision} & \textbf{Recall} & \textbf{F1-Score} \\
\hline
Benign & 1.00 & 1.00 & 1.00 \\
\hline
DoS & 0.98 & 1.00 & 0.99 \\
\hline
Probe & 0.99 & 1.00 & 1.00 \\
\hline
DDoS & 1.00 & 1.00 & 1.00 \\
\hline
Brute Force & 1.00 & 1.00 & 1.00 \\
\hline

Web Attack & 0.99 & 0.97 & 0.98 \\
\hline
Botnet & \textbf{0.87} & \textbf{0.46} & \textbf{0.60} \\
\hline
Infiltration & 1.00 & \textbf{0.67} & \textbf{0.80} \\
\hline

\end{tabular}
\label{tab:classification-metrics}
\end{table}

Table \ref{tab:classification-metrics} offers further elaboration on the performance of the IDS (the RF model), with metrics such as Precision, Recall, and F1-Score. It can be observed the IDS delivered high classification performance for six out of the eight classes. The two classes with lower classification performance are Botnet and Infiltration, with precision, recall, and F1-score values of 0.87, 0.46, 0.60 and 1.00, 0.67, 0.80, respectively. From Table \ref{tab:newclass}, it is observed that the instances for these two classes are significant lesser than most of the other classes. This highlights a significant challenge within the field of machine learning: the scarcity of available data. This scarcity of data can pose a significant obstacle to achieving optimal model performance. For the Infiltration class, there may still be room for improvement in both recall and F1-score. However, as previously mentioned, our strategy involves utilizing GAN models to generate additional instances based on original samples to enhance the IDS. Given that the Infiltration class contains only 36 instances, when the dataset (CIC-IDS2017) has 78 features   \cite{zhou2020building}, therefore our ability to enhance performance is constrained by the limited amount of available data, as it poses a serious challenge (At least a few hundred Infiltration samples which are several multiples of the number of features may be needed to generate realistic attack samples.) for GAN models to be able to effectively acquire the knowledge required to generate convincing and realistic samples.

On the other hand, the Botnet class consists of 1956 instances, presents a promising opportunity for enhancing its classification performance. As a result, our primary focus is directed toward this class, and we intend to harness GAN models to generate additional synthetic instances exclusively within this category.

\subsection{The Basic GANs}

A GAN consists of two neural networks, a generator and a discriminator. The generator network is responsible for producing new data instances that mimic a provided dataset.  The discriminator network evaluates whether a given data instance belongs to the authentic dataset (real) or if it was generated by the generator (fake). 

During the training process of a GAN, a competitive minimax game occurs between the generator and discriminator networks. The generator tries to produce data that is increasingly more realistic to deceive the discriminator, while the discriminator aims to improve its ability to differentiate between real and fake data. As a result of this adversarial process, the generator generates data that becomes progressively more convincing over time and ultimately it is almost indistinguishable from genuine data.

\subsubsection{Vanilla GAN (GAN)}

A GAN model that utilizes binary cross-entropy as its loss function is commonly known as a Vanilla GAN \cite {dunmore2023comprehensive} or simply a GAN. In this model, the discriminator is tasked with discerning between real and generated data. It produces an output \(D(x)\), which represents the probability that the input \(x\) is real (as opposed to generated). The discriminator's loss function is often represented as follows:

\begin{equation}
L_D = -\mathbb{E}_{x \sim p_{\text{data}}(x)} [\log D(x)] - \mathbb{E}_{x \sim p_{\text{gen}}(x)} [\log(1 - D(x))]
\end{equation}

 \(-\mathbb{E}_{x \sim p_{\text{data}}(x)} [\log D(x)]\) represents the expected value of the logarithm of the discriminator's output for real data. It encourages the discriminator to assign high probabilities (close to 1) to real data. While \(-\mathbb{E}_{x \sim p_{\text{gen}}(x)} [\log(1 - D(x))]\) represents the expected value of the logarithm of the complement of the discriminator's output for generated data. It encourages the discriminator to assign low probabilities (close to 0) to generated data.

The generator's objective is to minimize the probability that the discriminator correctly classifies the generated data as fake, which is essentially the flip side of the discriminator's objective. So, the generator's loss function is:

\begin{equation}
L_G = -\mathbb{E}_{x \sim p_{\text{gen}}(x)} [\log D(x)]
\end{equation}
The generator tries to minimize the expected value of the logarithm of the discriminator's output for generated data.

Where:
\begin{align*}
    L_D & = \text{Discriminator's loss} \\
    L_G & = \text{Generator's loss} \\
    x & = \text{Data samples} \\
    p_{\text{data}}(x) & = \text{Real data distribution} \\
    p_{\text{gen}}(x) & = \text{Generated data distribution} \\
    D(x) & = \text{Discriminator's output for data sample } x
\end{align*}

\subsubsection{Wasserstein GAN (WGAN)}

A WGAN \cite{arjovsky2017wasserstein} is a variant of traditional GANs that introduces a new loss function. In a standard GAN, the training process tries to minimize the binary cross-entropy loss for discriminator and generator networks. However, standard GANs can be challenging to train, as they suffer from problems like mode collapse and vanishing gradients.

Wasserstein GANs address some of these challenges by using the Wasserstein distance, alternatively recognized as the Earth-Mover's distance (EMD), serves as a loss function. This distance quantifies the lowest expense required to transform one probability distribution into another. 

The EMD between distributions \( P \) and \( Q \) is defined as:

\begin{equation}
\text{EMD}(P, Q) = \min \sum \sum d(i, j) \cdot f(i, j)
\end{equation}

Where:\\

$\text{EMD}(P,Q)$ = Earth-Mover's Distance between  $P$  and $Q$ \\ 

$d(i, j)$ = distance between data points $(i,j)$ from  $P$ and  $Q$ \\

$f(i, j)$ = amount of mass to be moved from  $i$ to  $j$ 

($i$ in  $P$ and $j$ in  $Q$) \\

Utilizing the Wasserstein distance as the loss function in GANs offers several advantages. First, it introduces continuous and smooth gradients into the training process, thereby enhancing training stability and mitigating issues like vanishing gradients. Moreover, the Wasserstein loss provides a more informative metric for training evaluation. By quantifying how closely the generated distribution aligns with the real data distribution, it equips the generator with a clearer objective: to produce samples that not only appear authentic but also capture the underlying data distribution faithfully. Additionally, Wasserstein GANs exhibit increased stability throughout the training process. This enhanced stability translates into faster convergence, ensuring that the model learns more efficiently.

\subsubsection{Wasserstein GAN with Gradient Penalty (WGAN-GP) }
A WGAN-GP model is a refinement of the WGAN model. WGAN-GP replaces weight clipping with a gradient penalty to enforce the Lipschitz constraint. This constraint helps in achieving a more stable training process and allows for meaningful Wasserstein distance computation. More details about WGAN-GP can be found in \cite{gulrajani2017improved}.

\subsection{The GAN Models and Settings in This Study}

In this study,  three distinct GAN models have been deployed for the purpose of generating new attack instances based on  the CIC-IDS2017 dataset. The initial two models comprise a Vanilla GAN and a WGAN, utilizing cross-entropy and Wasserstein distance respectively as their loss functions.  A specialized generative model designed specifically for tabular data, known as Conditional Tabular GAN (CTGAN) \cite{xu2019modeling}, constitutes the final model in the series.  This model excels in enabling conditional data generation, with a focus on maintaining the statistical characteristics and dependencies inherent in the original tabular dataset. In this model, the generator loss function is bases on  the Maximum Mean Discrepancy (MMD), which measures the distinctions between two distributions. While the discriminator loss is based on the principles of WGAN-GP, emphasizing the preservation of data quality and integrity during the generation process.

\subsubsection{Implementation and Settings for Vanilla GAN}

The generator comprises three fully connected layers, with the initial layer consisting of 25 neurons, accepting noise as input, applying the ReLU activation function, and initialized using the He uniform initializer.

The second layer has 50 neurons with ReLU activation. The last layer has as many units as there are features in the continuous scaled data (depends on the features used), and it uses sigmoid activation, which is typical for the output layer when generating data between 0 and 1.

The discriminator has also three  layers: The first layer comprises of 50 neurons, with ReLU activation, and takes the shape of the real data as the input. The second layer has 100 neurons and uses ReLU activation.
Then the final layer has 1 neuron and uses a sigmoid activation. It outputs a single value, which is interpreted as the probability that the input data is real (as opposed to generated).  %Note the the epochs used are 2000, batch size is 128 and and learning rate is 2e-4. 

\subsubsection{Implementation and Settings for WGAN}

The implemented WGAN model shared  similar settings as the Vanilla GAN in this study, except that it utilise the Wasserstein distance as its loss function.

\subsubsection{Implementation and Settings for CTGAN}

 We utilized and made modifications to a CTGAN model, which is originally based on the code available at \url{https://github.com/ydataai/ydata-synthetic}.  The generator has one input layer, three hidden layers and one output layer.    The noise and the label data concatenated together as the input.  The  first hidden layer  with neurons number equals to the input dimensionality (dim) and with ReLU as the activation. The second hidden layer  with dim * 2 neurons and ReLU activation and the third hidden layer with dim * 4 neurons and ReLU activation. The output layer generates synthetic data with the specified dimensionality (the output dimensionality).

The discriminator has also one input layer, three hidden layers and one output layer. Input data and label data are concatenated together as the input. The first hidden layer with dim * 4 neurons and ReLU activation. A dropout layer with a dropout rate of 0.1 is applied after the first hidden layer. The second hidden layer with dim * 2 neurons and ReLU activation. Another dropout layer with a dropout rate of 0.1 is applied after the second hidden layer. The third hidden layer with dim neurons and ReLU activation. The output layer with a single neuron and sigmoid activation. %Note the batch size used with this model is 500, epochs are 500 and learning rate is 2e-4.
%\clearpage

\subsection{Processing Botnet samples from CIC-IDS2017 and Generating New Samples}

Following the initial data processing, the original Botnet dataset consisted of 1956 samples. To enhance the generation of realistic network traffic samples using GAN models, we recognized the necessity to further divide and categorize these samples. Initially, we classified the original Botnet samples into two primary groups based on their destination ports: those linked to port 8080 and those associated with non-8080 port numbers. The choice of destination ports serves as an indicator of the application protocol, each of which possesses its distinct feature boundaries within network traffic. It's worth noting that a significant portion of the Botnet samples were associated with port 8080, which serves as an alternative port for the Hypertext Transfer Protocol (HTTP).

In further refining the approach, these two primary groups are further divided into smaller, more focused segments. The criterion for this division was straightforward: it is  observed that certain columns in the dataset predominantly contained just two or three distinct values. We tailored the dataset divisions to align with these observed patterns, creating smaller, more homogenous segments. These resulting smaller segments exhibited simplified data distributions. Subsequently, we employed the three GAN models to generate additional Botnet samples based on these refined, more homogenous segments.

\subsection{The Evaluation for Closeness}

Three ways in this study were employed to quantify the similarity (closeness) between the synthetic Botnet data and the Botnet data from the original CIC-IDS2017 dataset, more details will be provided in the following. 

\subsubsection{The Cosine Similarity }
Cosine similarity \cite{wang2024raman} is a metric employed to gauge the similarity between two non-zero vectors within an inner product space, and it finds relevance in our specific context. The closer its value is to 1, the greater the proximity between the two vectors. In fact, when the value equals 1, it signifies that the two vectors align perfectly in direction, indicating an absolute similarity.

  The generated samples exhibit a high level of similarity to the originals, indicating the effectiveness of these models (the three GAN models) in preserving the key characteristics of the original data. We have also observed that Vanilla GAN and WGAN models exhibited similar performance in terms of cosine similarity, and they  outperformed the CTGAN model.   In Table \ref{tab:CS}, we have provided 8 features (Flow\_Duration,  Total\_Length\_of\_Fwd\_Packets,  Flow\_Packets\_s,     Fwd\_ IAT\_Mean,    Bwd\_IAT\_Mean, \\  Fwd\_Packets\_s,    Packet\_Length\_Mean, and \\ Init\_Win\_bytes\_backward) and their corresponding  cosine similarity values between original Botnet samples and  generated samples by GAN, WGAN and CTGAN models. While there were more features available for selection, the primary reason we chose these specific 8 features is their high degree of representativeness.

\begin{table}
\centering
\caption{The Cosine Similarities Between Generated and Original  Botnet Instances for 8 Features.}
\label{tab:CS}
\begin{tabular}{| c | c | c | c | }
\hline
\textbf{Features} & \multicolumn{3}{ c |}{\textbf{GAN Models}}  \\ 
\cline{2-4}
& \textbf{Vallina GAN} & \textbf{WGAN} & \textbf{CTGAN} \\
\hline

Feature 1 & 0.8979 & 0.8986 & 0.8295 \\
\hline
Feature 2 & 0.7608 & 0.7607 & 0.5531 \\

\hline
Feature 3 & 0.9673 & 0.9653 & 0.9147 \\

\hline
Feature 4 & 0.9013 & 0.8977 & 0.8316 \\
\hline
Feature 5 & 0.8962 & 0.8956 & 0.8520 \\
\hline
Feature 6 & 0.9669 & 0.9655 & 0.8958 \\
\hline
Feature 7 & 0.8750 & 0.8750 & 0.7001 \\

\hline
Feature 8 & 0.9999 & 0.9999 & 0.9997 \\
\hline
\end{tabular}
\end{table}

\subsubsection{Cumulative sums}

Cumulative sums of a feature \cite{liu2023anomaly}, achieved by aggregating the values of a single feature, can also serve as a means to quantify the closeness between generated and initial samples. Upon comparing the cumulative sums of features between the generated samples (produced by Vallila GAN, WGAN, and CTGAN) and the original samples, we observed a consistent close alignment. Furthermore, our analysis led to the conclusion that Vallila GAN and WGAN models exhibited similar performance in terms of cumulative feature sums, surpassing the performance of CTGAN.

 \begin{figure}[!ht]
  \includegraphics[width=1.0 \linewidth]{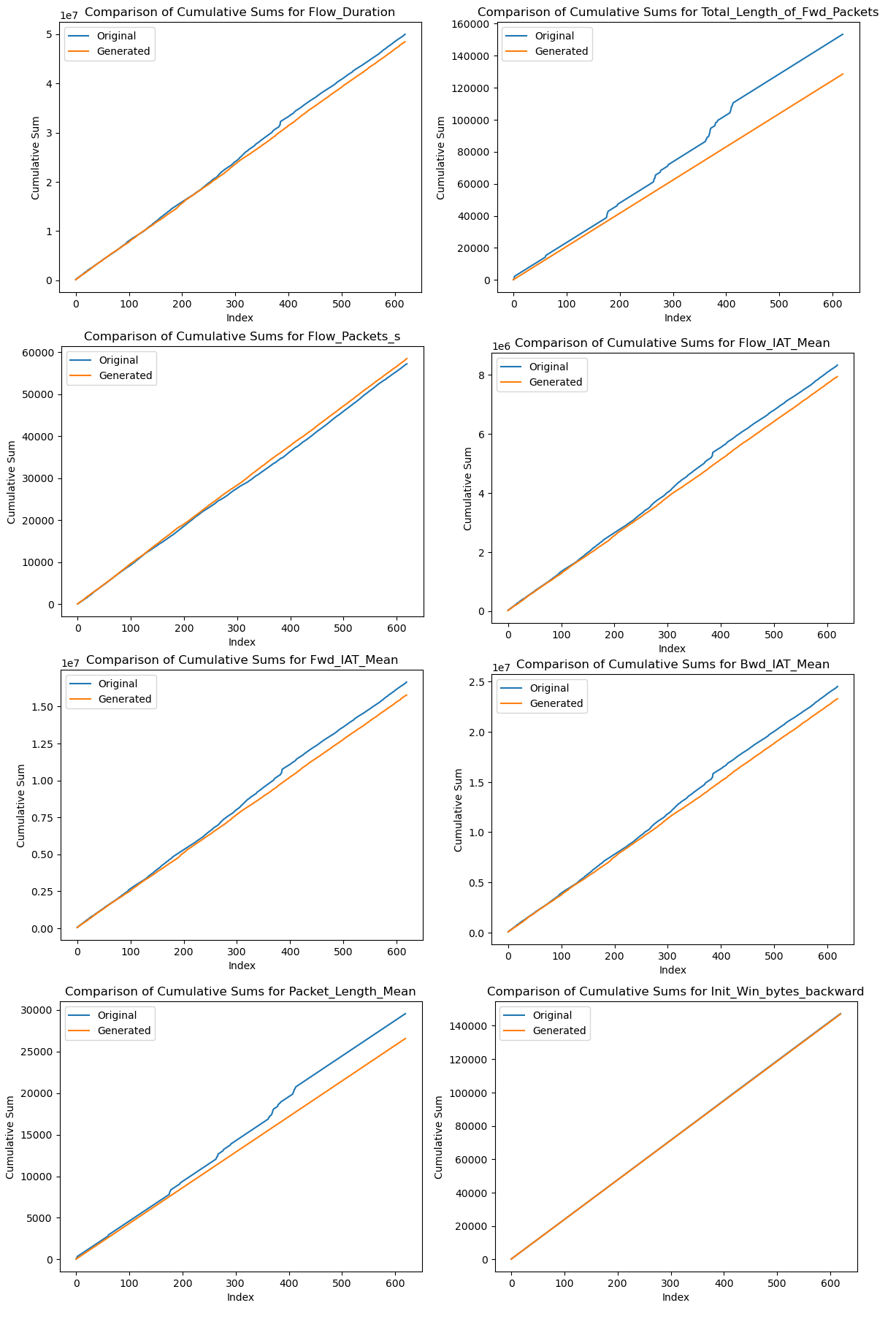}
  \caption{The cumulative sums for 8 features for GAN generated and original Botnet samples.}
  \label{fig:Gansum}
\end{figure}

 \begin{figure}[!ht]
  \includegraphics[width=1.0 \linewidth]{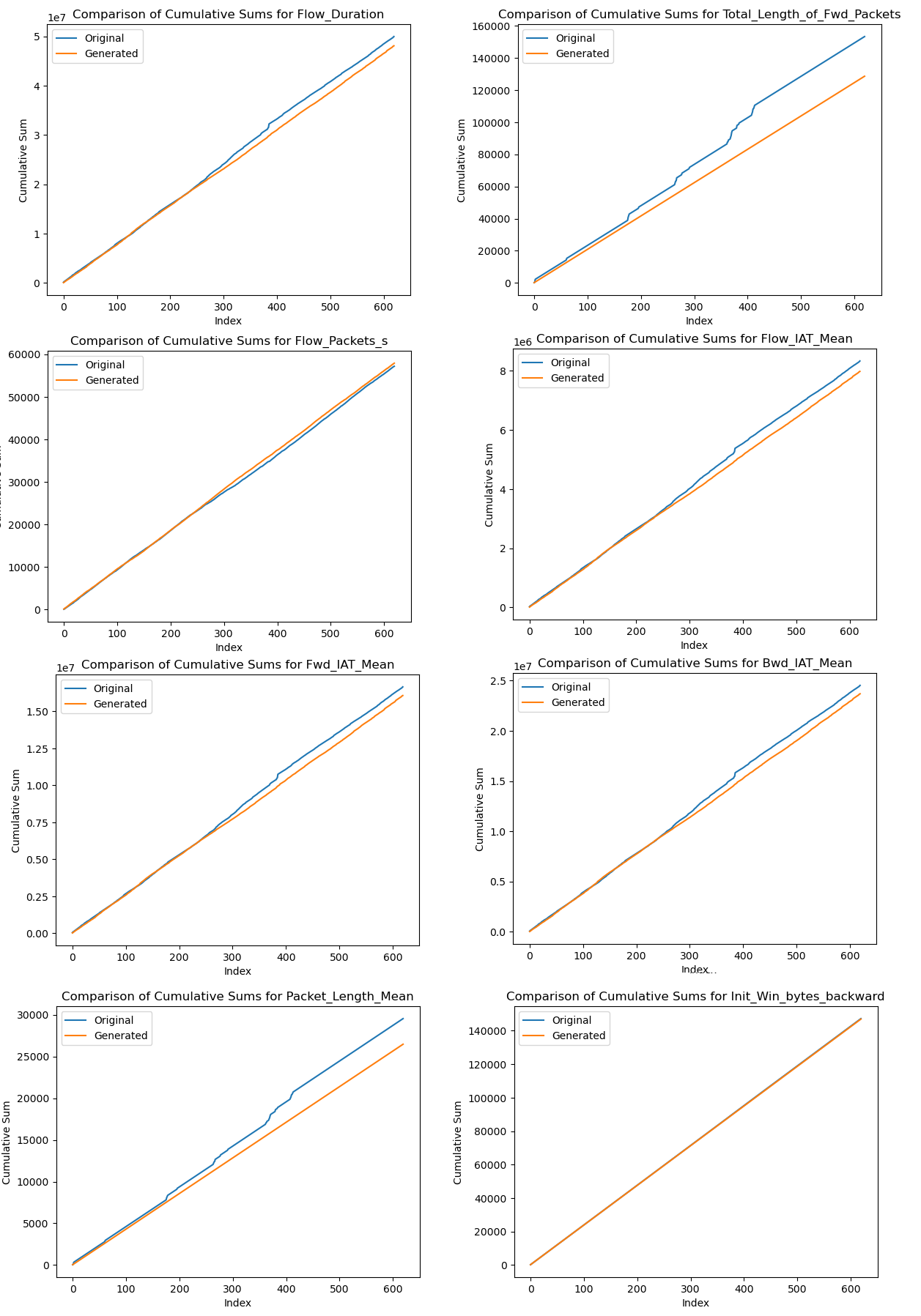}
  \caption{The cumulative sums for 8 features for WGAN generated and original Botnet samples.}
  \label{fig:Wgansum}
\end{figure}

 \begin{figure}[!ht]
  \includegraphics[width=1.0 \linewidth]{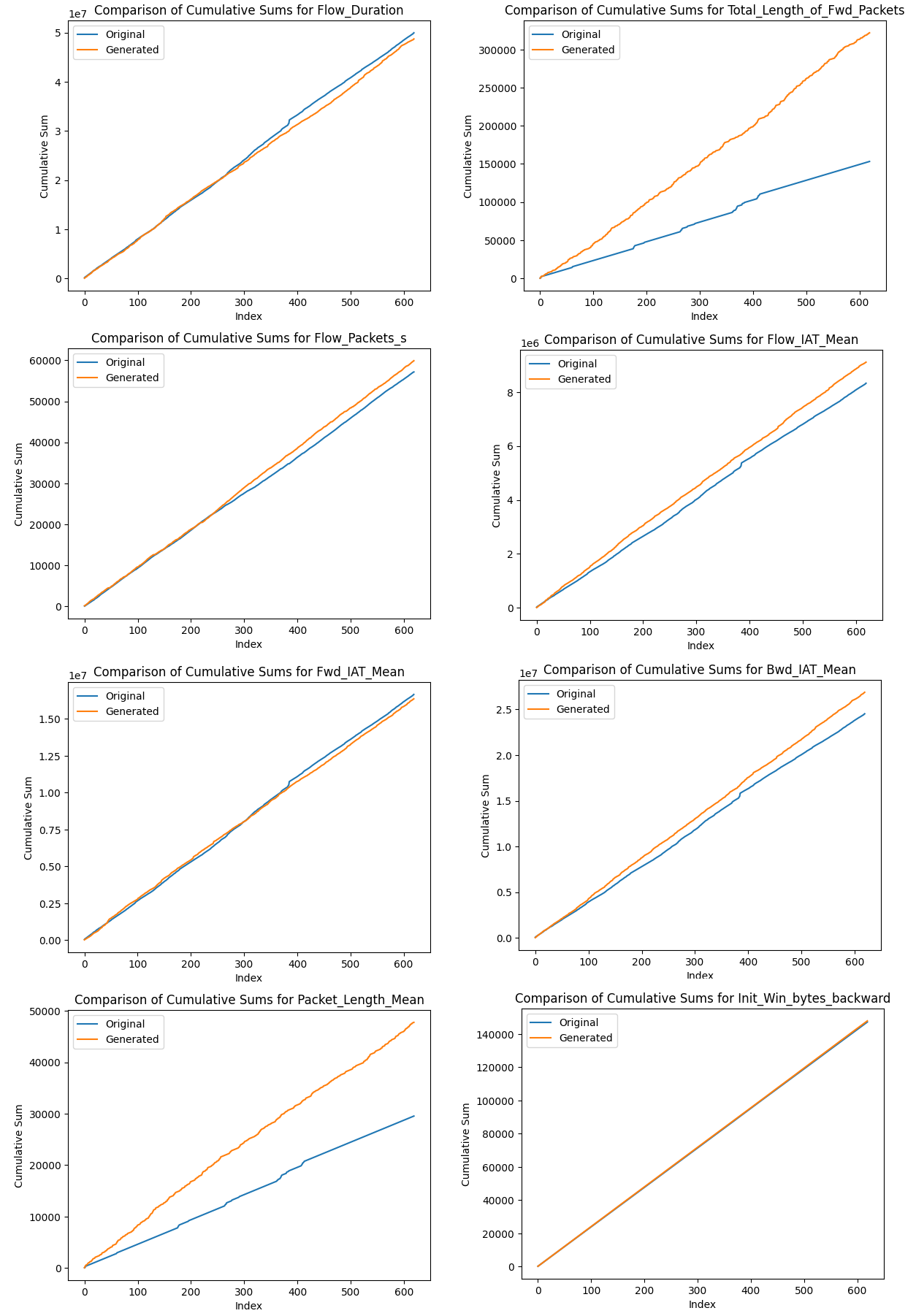}
  \caption{The cumulative sums for 8 features for CTGAN generated and original Botnet samples.}
  \label{fig:CTgansum}
\end{figure}

Figures \ref{fig:Gansum}  to \ref{fig:CTgansum} depict cumulative sums for the 8 features we mentioned in the previous subsection (These are the same eight features as presented in Table \ref{tab:CS}) within three distinct groups.The blue lines are cumulative sums for samples from the original Botnet samples while the orange lines are cumulative sums for generated samples. We can observe that, for WGAN and Vanilla GAN, the cumulative sums of the eight selected features closely match between the generated samples and the original Botnet data. However, when considering CTGAN, there is a significant divergence in the cumulative sums of certain features in comparison to the original samples.

\subsubsection{Validating with ML Algorithms}

Another approach in assessing the closeness between generated samples and the original samples is to employ machine learning algorithms. These algorithms (such as Random Forest and Decision Tree) can provide valuable insights into the quality and fidelity of the generated data relative to the initial dataset.

We utilized GAN, WGAN, and CTGAN to generate 1,956 synthetic Botnet samples, mirroring our reference dataset with the same number of original Botnet instances. Additionally, we extracted two groups of 10,000 Benign samples from the CIC-IDS2017 dataset.

In the first experiment, we combined 1,956 CTGAN-generated samples with 10,000 Benign samples to form a dataset (dataset 1). We also created another dataset  by combining the original 1,956 Botnet samples with another set of 10,000 Benign samples (testing set). A random forest model trained on 80\% of dataset 1 achieved a precision, recall, and F-score of 1.00. When tested on the testing set, it maintained strong performance with a precision of 0.98, recall of 0.94, and an F-score of 0.96.

 The second experiment involved using 1,956 Vanilla GAN-generated samples combined with the 10,000 Benign samples (dataset 2). We applied the RF model with strong results: a precision of 1.00, a recall of 0.99, and an F-scores of 1.00 on the remaining 20\% of dataset 2. Applying the model on the testing set yielded a precision of 0.99, recall of 0.92, and an F-score of 0.95.

In the third experiment, we employed 1,956 WGAN-generated samples combined with the 10,000 Benign samples(dataset 3). The RF model achieved  all measure metrics of 1.00 when tested on the remaining 20\% of dataset 3. Applying the model on the testing set resulted in a precision of 0.99, recall of 0.96, and an F-score of 0.98.

Similar results were obtained using the Decision Tree classifier. These experiments underscore the consistency of the samples generated by the three GAN models, indicating a high level of resemblance to the original samples.

\section{Intrusion Detection Enhancement with More Botnet
Samples Generated} \label{enhancing}

As we mentioned, we generate samples based on the segments divided from the original Botnet samples.  For example, if we want to generate 1956 X 4 (4 times) Botnet samples with WGAN model. We will use the WGAN model to generate proportionally more samples based on the original divided segments. In this particular scenario, each segment of the samples will be generated four times in size and subsequently assembled together. 

 We leveraged the three GAN models—WGAN, Vanilla GAN, and CTGAN—to generate augmented sets of Botnet samples. Each set comprised 1,956 samples, multiplied by 4, 49, and 99, respectively, were categorized into distinct groups for the IDS enhancement  evaluation.

\subsection{Enhance with WGAN-Generated Samples} \label{sec:WGAN}

For the set which generated by the WGAN model samples, we separated these sample into three groups and evaluated them separately.

In the group I,  we replaced the original 1,956  Botnet samples in the processed CIC-IDS2017 dataset with four times the originals (1956X4) of Botnet samples generated by the WGAN model. Using the same RF model as in the IDS baseline (as show in section \ref{sec:Baseline}, 8:2 ratio for training and testing), we achieved strong Botnet classification results, with precision, recall, and F-score all exceeding 0.97. Testing the RF model on the original 1,956 Botnet samples resulted in a precision, recall and F-score of 1.00,  0.74 and 0.85, respectively.

In group II,  here, we augmented the dataset with 49 times the original number of samples, again replacing the 1,956 original Botnet samples in the   processed CIC-IDS2017 datase.The RF model exhibited good performance on the generated samples, with all metrics reaching 1.00. When tested on the original 1,956 Botnet samples, the RF model achieved a precision, recall and F-score  of 1.00, 0.76 and 0.87, respectively.

In group III, we employed 99 times the original number of samples, again replacing the 1,956 original Botnet samples in the processed CIC-IDS2017 datase. The RF model consistently delivered good results for Botnet classification, with all metrics reaching 1.00. Testing this model on the original 1,956 Botnet samples yielded a precision, recall and F-score  of 1.00, 0.82 and  0.90, respectively.
More details can be found in   Tables \ref{tab:499Gen} and  \ref{tab:499}.

\subsection{Enhance with Other Two GAN-Generated Samples}\label{sec:VGAN}

 We conducted parallel experiments with Vanilla GAN and CTGAN generated samples following a similar protocol as the WGAN experiments.  Performance metrics for Botnet classification were evaluated for each group, both on the generated samples and the original 1,956 Botnet samples. Additional details can be found in Tables \ref{tab:499Gen} and \ref{tab:499}.

\begin{table}[htbp]
\centering
\caption{Classification Performance of the IDS on Generated Botnet Samples.}
\begin{tabular}{|l|c|c|c|}
\hline
\textbf{Model (Numbers)} & \textbf{Precision} & \textbf{Recall} & \textbf{F1-Score} \\
\hline
WGAN(4 times) & 0.97 & 1.00 & 0.98 \\
\hline
WGAN(49 times)  & \textbf{1.00} & \textbf{1.00} & \textbf{1.00} \\
\hline
WGAN(99 times)  & \textbf{1.00} & \textbf{1.00} & \textbf{1.00} \\
\hline

Vanilla GAN (4 times) & 0.98 & 0.99 & 0.98 \\
\hline

Vanilla GAN (49 times) & \textbf{1.00} & \textbf{1.00} & \textbf{1.00} \\
\hline

Vanilla GAN (99 times) & \textbf{1.00} & \textbf{1.00} & \textbf{1.00} \\
\hline

CTGAN (4 times) & 0.99 & 0.68 & 0.81 \\
\hline

CTGAN (49 times) & 0.97 & 0.92 & 0.95 \\
\hline

CTGAN (99 times) & 0.95 & 0.99 & 0.97 \\

\hline
\end{tabular}
\label{tab:499Gen}
\end{table}

\begin{table}[htbp]
\centering
\caption{Trained IDS Performance on Original Botnet Samples.}
\begin{tabular}{|l|c|c|c|}
\hline
\textbf{Model (Numbers)} & \textbf{Precision} & \textbf{Recall} & \textbf{F1-Score} \\
\hline
WGAN(4 times) & 1.00 & 0.74 & 0.85 \\
\hline
WGAN(49 times)  & 1.00 & 0.76 & 0.87 \\
\hline
WGAN(99 times)  & \textbf{1.00} & \textbf{0.82} & \textbf{0.90} \\
\hline

Vanilla GAN (4 times) & 1.00 & 0.66 & 0.80 \\
\hline

Vanilla GAN (49 times) & 1.00 & 0.76 & 0.87 \\
\hline

Vanilla GAN (99 times) & 1.00 & 0.81 & 0.90 \\
\hline

CTGAN (4 times) & 1.00 & 0.43 & 0.60 \\
\hline

CTGAN (49 times) & 1.00 & 0.76 & 0.86 \\
\hline

CTGAN (99 times) & 1.00 & 0.77 & 0.87 \\

\hline
\end{tabular}
\label{tab:499}
\end{table}

\section{Discussion} \label{Discussion}

Based on the analysis of the  preceding subsections, it is evident that increasing the number of generated Botnet samples in training, results in a significant performance improvement for classifying these generated Botnet instances. 

These experiments aimed to demonstrate the potential of augmenting the dataset with GAN-generated Botnet samples to enhance intrusion detection. The results showcase the increased volume of data and its impact on model performance, with each GAN model offering distinct advantages.

For instance, when using both WGAN and Vanilla GAN and augmenting the dataset with forty-nine and ninety-nine times the generated samples, the IDS achieved flawless classification, with precision, recall, and F1-scores all reaching the maximum value of 1. This trend can be observed from Table \ref{tab:499Gen}, and note that in these cases, the IDS was trained using augmented datasets and subsequently tested on the generated datasets.

  More importantly, with the generation of a larger volume of Botnet samples using the three GAN models, the IDS demonstrated enhanced training effectiveness.   Table \ref{tab:499} presents the classification results for the original 1956 Botnet samples, achieved by training the IDS with augmented datasets of more generated Botnet samples using three different GAN models. For example, upon incorporating ninety-nine Botnet samples (WGAN generated) into the training dataset, the IDS exhibited considerable  performance improvements. These improvements were evident in the precision score, which reached a perfect 1.00 (a notable 13\% enhancement), a recall of 0.81 (a substantial 35\% improvement), and an F1-score of 0.90 (a substantial 30\% improvement) when classifying the original set of 1956 Botnet samples. This represents a significant leap from the baseline results, where precision was at 0.87, recall at 0.46, and the F1-score at 0.60, as the results clearly indicate. These findings underscore the substantial progress in the IDS's training, enabling it to successfully detect previously elusive samples. 

 Another intriguing observation is that, in the cases of both the WGAN and Vanilla GAN models, the IDS experiences substantial improvement even when only four times the number of samples are generated for training, as compared to the IDS baseline. However, it's noteworthy that when we increased the volume to forty-nine and ninety-nine times, the IDS's performance demonstrated further enhancements, although these improvements were not as substantial as the initial fourfold increase. This trend could be attributed to the diminishing returns associated with the increased volume of generated data, where the initial increase yields more significant gains in performance compared to subsequent increments.

Concerning the samples generated by CTGAN models, a noticeable trend in IDS enhancement was also evident. However, it is worth noting that these enhancements are comparatively less pronounced when compared to the improvements seen with the samples generated by the other two GAN models.  It's worth highlighting that when using only four times the number of original samples (1956X4), the improvement in IDS classification remains rather limited when compared to the baseline. This could because CTGAN generated samples have less closeness or similarities (implied in Table \ref{tab:CS} and Figure \ref{fig:CTgansum}) to the original Botnet samples from the CIC-IDS 2017, as compared to the WGAN and Vanilla GAN models.  Notably, when we increased the volume to forty-nine times, the IDS exhibited significant enhancements. However, it's interesting to observe that the incremental benefit of generating ninety-nine times the samples over the forty-nine times case was almost negligible.

The reduced efficacy (for IDS performance enhancement) of the Botnet samples generated by CTGAN models, in contrast to the more effective results observed with WGAN and Vanilla GAN models, can likely be attributed to the specific approach employed in the generation process. In our approach, the original Botnet samples were divided into smaller segments, and new attack samples were subsequently generated proportionally based on these fragmented components. Within each of these smaller segments, it was common to encounter many columns with consistent, singular values. These  single values within the segments contributes to the simplification of their distributions. The increased sophistication of CTGAN model, in comparison to WGAN and Vanilla GAN models, does not inherently guarantee its effectiveness in handling simpler distributions when contrasted with the capabilities of WGAN and Vanilla GAN models.

We would like to underscore an important point: as we augment our dataset by generating a larger volume of samples and incorporating them into the original dataset (CIC-IDS2017) for training our IDS, its performance in identifying classes other than Botnet exhibits consistent stability. Specifically, the Web Attack and Infiltration classes are the only ones showing a marginal impact, with any enhancements falling within a 4\% range. This stable performance holds true across all the scenarios involving different GAN models and varying numbers of Botnet samples used during IDS training. However, the marginal improvements observed in these two classes may be more attributable to the effects of an increased number of Botnet samples rather than genuine performance enhancements. Table \ref{tab:classificationWGANX99}.
shows the classification performance    when we generated 99 times the original 1956 Botnet samples and replaced them  with the original 1956 samples in the dataset to train the IDS. Comparing these results to those in Table \ref{tab:classification-metrics}, minor variations in the Recall and F1-score metrics are noticeable for the Infiltration class, alongside a slight modification in the Recall value for the Web Attack class. This observation holds significant importance because if the IDS classification performance for other classes had been impacted significantly during the process, it could have raised questions about the validity of our research methodology. However, since we have confirmed that the IDS's performance for other classes remained stable, it reinforces our confidence in the feasibility of creating more precise samples tailored to the Botnet class. This, in turn, opens up an exciting opportunity to enhance the IDS's performance specifically for this class of cyber threats.

\begin{table}[htbp]
\centering
\caption{Classification performance with 1956X99 WGAN Generated Botnet Samples.}
\begin{tabular}{|l|c|c|c|}
\hline
\textbf{Class} & \textbf{Precision} & \textbf{Recall} & \textbf{F1-Score} \\
\hline
Benign & 1.00 & 1.00 & 1.00 \\
\hline
DoS & 0.98 & 1.00 & 0.99 \\
\hline
Probe & 0.99 & 1.00 & 1.00 \\
\hline

DDoS & 1.00 & 1.00 & 1.00 \\
\hline
Brute Force & 1.00 & 1.00 & 1.00 \\
\hline
Web Attack & 0.99 & \textbf{0.98} & 0.98 \\
\hline
Botnet & 1.00 & 1.00
 &1.00 \\
\hline
Infiltration & 1.00 & \textbf{0.71} & \textbf{0.83} \\
\hline

\end{tabular}
\label{tab:classificationWGANX99}
\end{table}

In the current research, an RF model was utilized as the IDS, training it on the GAN-augmented CIC-IDS2017 dataset. This approach yielded outstanding results, with a precision of 1.00, a recall of 0.82, and an F1-score of 0.90 when applied to the original Botnet samples. These results signify a noteworthy advancement and, to the best of our knowledge, establish a new benchmark as the current best-reported performance in the literature. For example,  Keserwani et al. \cite{keserwani2021smart}    harnessed both Grey Wolf Optimization and Particle Swarm Optimization to extract features from the CIC-IDS2017 dataset. These extracted features were then fed into an RF classifier, resulting with a precision of 1.00, a recall of 0.60, and an F1-score of 0.75 for Botnet classification.  Lee et al. \cite{lee2021gan} employed a GAN model to generate 10,000 Botnet samples, which were integrated into the CIC-IDS2017 dataset. Using an RF as their IDS, they reached a precision of 0.86, a recall of 0.53, and an F1-score of 0.66 for Botnet classification.

\section{Conclusion} \label{Conclusion}

To address the critical challenge of data scarcity in NIDS training datasets, we have introduced a novel approach that integrates GANs into the NIDS framework. Our approach involves the development and implementation of three distinct GAN models that generate synthetic network traffic data, closely mimicking real-world network behavior while targeting specific predefined anomalous activities. Through extensive experimentation on a benchmark dataset (CIC-IDS2017) and generated datasets, we have demonstrated the efficacy of our proposed approach in NIDS classification enhancement. Our findings indicate that the integration of GANs into NIDS systems can lead to enhancements in intrusion detection performance, enabling the detection of previously undetected intrusion attempts. This research contributes to the continuous evolution of NIDS and shows the importance of adapting to the ever-changing landscape of cyber threats. As cyberattacks become increasingly sophisticated, the incorporation of GANs in NIDS represents a proactive and effective strategy for enhancing the security of digital infrastructures, thereby fortifying organizations against potential intrusions and data breaches.

\printcredits

%\bibliographystyle{cas-model2-names}
\bibliographystyle{elsarticle-num}
\bibliography{cas-refs}


%\bio{Figs/pic1}


\end{document}